\RequirePackage{iftex}
\ifPDFTeX\pdfoutput=1\fi
\documentclass[11pt]{article}

\usepackage[margin=1in]{geometry}
\usepackage{amsmath}
\usepackage{amssymb}
\usepackage{booktabs}
\usepackage{enumitem}
\usepackage{array}
\usepackage{tabularx}
\usepackage[section]{placeins}
\usepackage{flafter}
\usepackage{microtype}
\usepackage{tikz}
\usepackage{url}
\ifXeTeX
  \usepackage[xetex,hidelinks]{hyperref}
\else
  \usepackage[hidelinks]{hyperref}
\fi
\usetikzlibrary{arrows.meta,backgrounds,fit,positioning,shapes.geometric}
\newcolumntype{L}[1]{>{\raggedright\arraybackslash}p{#1}}
\newcolumntype{Y}{>{\raggedright\arraybackslash}X}
\title{XScientist: A Git-Like Research Protocol for Long-Running Autonomous Scientific Discovery}

\author{
Jixiang Luo\\
\texttt{jixiangluo85@gmail.com}
}

\date{}

\begin{document}
\maketitle

\begin{center}
\small Repository: \url{https://github.com/smileformylove/XScientist}
\end{center}

\begin{abstract}
Autonomous research systems can generate plausible papers while losing the
decisions, failed branches, and evidence needed to inspect or continue the
work. We present XScientist, a local-first, git-like protocol that treats
research state, rather than a manuscript, as the unit of continuation.
Hypotheses, experiment attempts, observations, claims, reviews, and handoffs
are represented as typed, content-addressed objects in an exploration graph.
Immutable checkpoints, explicit negative outcomes, claim--evidence closure,
replay boundaries, and authority-aware gates make each transition inspectable
without treating a passing integrity check as scientific truth. The protocol
exports a portable Agent-Native Research Artifact (ARA) that another agent or
human can inspect, fork, verify, and extend. A reference implementation
integrates planning, execution, review, repair, and supervised long-running
operation while preserving provenance across these stages. We evaluate the
protocol with controlled artifact-integrity workloads and matched external task
pilots, keeping native task performance separate from evidence and audit
claims. The result is an interoperability and accountability layer for
long-running autonomous science, with explicit boundaries where human
judgment and independent evaluation remain necessary.
\end{abstract}

\section{Introduction}

Large language models now automate substantial parts of a scientific workflow:
ideation, experiment scripting, result interpretation, manuscript drafting,
and review simulation \cite{ai-scientist,ai-scientist-v2}. The difficult case is
not a single successful run. It is a research program that branches, fails,
changes direction, and is resumed by another agent or a human weeks later.
For that setting, a final PDF is an impoverished interface: it hides which
decisions were tried, which evidence was discarded, and which claims still
depend on an unverified step.

The systems question is how to maintain executable, inspectable research state
across those transitions. Experiment tracking and provenance standards already
preserve substantial execution context \cite{mlflow,datalad,workflow-run-ro-crate};
research agents also retain search trees and intermediate findings
\cite{ai-scientist-v2,kosmos}. The unresolved design problem is how these records
govern the next research decision. If a mutable run directory is the only
reference, a correction can obscure its predecessor. If claims are detached
from evidence identities, later edits can invalidate their support silently.
If failed attempts are omitted, a successor must reconstruct why alternatives
were abandoned. These failure modes motivate a common contract for research
objects, transitions, and evidence requirements.

\subsection{Design Principles}
XScientist addresses this problem with a git-like research protocol. The
system is designed around four principles:

\begin{enumerate}[leftmargin=*]
  \item Each research stage should emit structured artifacts that can be
  inspected without replaying the entire workflow.
  \item Research state should support continuation from a specific experiment
  node instead of requiring reconstruction from the final PDF.
  \item Manuscript assertions should remain linked to the experiment nodes
  and evidence artifacts that support them.
  \item Failed branches, reviewer objections, integrity warnings, and blocked
  gates should remain first-class outputs rather than hidden pipeline noise.
\end{enumerate}

The resulting system can run as a local project runner, a continuous paper
generator, or a daemon, but its primary output is not prose. It is a typed,
content-addressed history in which hypotheses, experiment attempts,
observations, claims, reviews, and handoffs form an exploration graph. A
portable Agent-Native Research Artifact (ARA) exports that history for another
agent or reviewer.

\subsection{Contributions}
Our central thesis is that long-running autonomous science should promote
research state, rather than a manuscript, as the unit of continuation and
audit. We make three contributions:

\begin{enumerate}[leftmargin=*]
  \item Research VCS gives scientific objects typed transitions, immutable
  checkpoints, explicit negative results, and claim--evidence closure rules
  that ordinary file history does not provide.
  \item ARA packages the exploration graph, provenance, executable inputs,
  and diagnostic outcomes so that a downstream agent can fork a recorded node
  without reverse-engineering a final paper.
  \item On real computational artifacts, we compare protocol checks with
  ordinary Git, complete byte hashes, and normalized JSON hashing under a
  fixed corruption matrix. The study measures auditability and replay
  boundaries; it does not claim superior scientific discovery or a higher
  benchmark score.
\end{enumerate}

This framing determines how we evaluate the system. The prespecified endpoints
are per-fault rejection counts and intact-control false rejections. A post hoc
comparison examines storage baselines and the presence of diagnostic context.
Published autonomous-science benchmark scores
remain separate endpoints that require matched agents, budgets, and independent
evaluators; no such score is inferred from the artifact study.

This separation is necessary because autonomous-science systems are evaluated
at several incompatible levels. A paper-generation study may report reviewer
ratings, a code-search system may report task success, and a laboratory system
may report human judgments or domain measurements. Even when two systems use
the same benchmark, model version, tool access, search budget, validation
protocol, intervention policy, and treatment of failed or timed-out runs can
differ. A single aggregate score therefore risks measuring the evaluation
protocol rather than the research capability. We use matched task pilots only
where the task contract and official evaluator can be held fixed, and keep
artifact integrity, scientific-task performance, and human certification as
separate claims.

\section{Related Work}

\subsection{Autonomous-science systems and evaluation}
The AI Scientist introduced an end-to-end loop for automated idea generation,
experiment execution, paper writing, and review \cite{ai-scientist}; related
open research prototypes such as autoresearch also explore lightweight
automation loops for scientific work \cite{autoresearch}. AI
Scientist-v2 extends this line with agentic tree search and workshop-level
paper generation \cite{ai-scientist-v2}. XScientist follows the same broad
goal of automating research workflows, but shifts the main abstraction from
``generate a paper'' to ``produce a forkable research protocol artifact.''
The protocol layer is intended to make continuation, review, and
cross-system handoff explicit. Parts of the execution, tree-search, writing,
and review runtime are modified derivatives of AI-Scientist-v2, with transitive
AIDE lineage. These components are inherited foundations, not new algorithms
introduced by this paper; the additions described here concern research-state
semantics, artifact exchange, and governance.

The published systems report different endpoints and should not be collapsed
into one leaderboard. The AI Scientist reports 38 completed diffusion papers
from 51 ideas with Claude Sonnet 3.5, with a mean automated reviewer score of
3.82 on the NeurIPS 2--6 scale (maximum 6.0); its separate calibrated GPT-4o
reviewer benchmark uses 500 ICLR 2022 papers and reports 0.65 +/- 0.04
balanced accuracy (95\% bootstrap CI) \cite{ai-scientist}. AI Scientist-v2 replaces
the fixed experiment template with a staged best-first tree search. In its
reported ICLR 2025 ICBINB workshop study with 43 total submissions, three
selected AI-generated manuscripts received one acceptance decision; the
accepted manuscript's three-reviewer mean was 6.33/10 (6, 6, 7), but it was
withdrawn before publication. This is not an unbiased 1/3 autonomous success
rate \cite{ai-scientist-v2}. AIDE treats machine-learning engineering
as search over executable code and reports performance above the human median
on 8/16 Weco-Kaggle Lite tasks; its Table 3 reproduction of MLE-bench reports
16.9 +/- 1.1 percentage points any-medal pass@1 for AIDE with o1-preview over
75 competitions (mean +/- one standard error over 16 seeds)
\cite{aide,mle-bench}.

Agent Laboratory adds role-specialized literature review, experimentation,
report writing, and optional human checkpoints. Its autonomous quality study
contains 15 papers; the reported human experimental-quality means are 2.6/5
(gpt-4o), 3.2/5 (o1-mini), and 2.9/5 (o1-preview)
\cite{agent-laboratory}. AgentRxiv extends this workflow with a shared archive
of prior reports. In its MATH-500 study, one sequential laboratory uses
gpt-4o-mini for experiments and 40 generations, improving a 70.2\% baseline to
the best observed 78.2\% while retrieving five prior AgentRxiv papers and five
arXiv papers; three parallel laboratories (40 generations each) report 79.8\%
best accuracy
\cite{agentrxiv}. These results concern generated research quality or task
accuracy, whereas this paper evaluates whether stored research artifacts
remain checkable after corruption and replay. We therefore report the
published numbers for context, not as head-to-head evidence, and do not claim
to outperform these systems.

The comparison is intentionally separated from a discovery leaderboard. We use
four axes: end-to-end scientific-task outcome, validation-layer ablations on a
fixed run, native artifact-audit evidence, and long-horizon recovery. The cited
systems report results mainly on the first axis. XScientist evaluates the
second and third axes on fixed artifacts and does not evaluate recovery.
``NR in the cited version'' means that a paper does not report that endpoint;
it is not evidence that the system cannot support it. A future common
executable-task adapter could compare open implementations under matched
models, seeds, inputs, and budgets, but that would not be a shared
scientific-discovery score.

Two newer systems broaden the comparison beyond code and report generation.
Co-Scientist uses a multi-agent hypothesis-generation and tournament-evolution
loop with human research objectives and biomedical validation; its Nature report
describes preliminary cell-viability inhibition for three of five expert-selected
candidate compounds, screened using four AML cell lines and a TK6 control,
with three biologically independent dose-response experiments
\cite{co-scientist}. Kosmos maintains a structured world model of research
summaries, with report statements and figures linked to publications or
Jupyter notebooks. Expert assessment of 102 statements from three representative
reports found 79.4\% statement accuracy
\cite{kosmos}. These are domain outcomes and claim-audit measurements, not
artifact-integrity benchmarks. They further motivate XScientist's separation
between a durable research record and a scientific claim that still requires
domain review.

Other recent proposals broaden the design space with chain-of-evidence
records, human--agent feedback, progressive discovery, autonomous innovation,
mathematical discovery, or reflective search, including ScientistOne,
AutoResearchClaw, DeepScientist, AI-Researcher, FAR, and MARS
\cite{scientistone,autoresearchclaw,deepscientist,ai-researcher,far,mars}.
We cite these systems as adjacent designs; none is rerun or scored in the
present artifact-integrity study.

The comparison is therefore made on explicit axes rather than on a single
leaderboard: end-to-end scientific-task outcome, continuity of research state,
claim--evidence grounding, replay and recovery, and human handoff. The cited
systems provide useful task execution, search, review, or memory mechanisms,
but their reported artifacts do not define a common protocol for all five
axes. No external autonomous agent is rerun here. The matched quantitative
comparison in Section~\ref{sec:evaluation} uses the same XScientist artifacts
and mutations for protocol checks and storage baselines; the source-audited
capability inventory and benchmark status are provided in the repository review
archive and in Appendix~\ref{app:benchmarks}.

\subsection{Agent execution, tracking, and provenance}
ReAct combines reasoning and acting in language models \cite{react};
Reflexion studies verbal feedback loops for language agents
\cite{reflexion}; Tree of Thoughts explores deliberate branching search
\cite{tree-of-thoughts}; Voyager demonstrates open-ended skill accumulation
in an embodied environment \cite{voyager}. Software-engineering agents such
as SWE-agent show how interface design affects autonomous repair
\cite{swe-agent}; AIDE explores code-space search for ML engineering tasks
\cite{aide}, and AutoGen supports multi-agent orchestration \cite{autogen}.
XScientist applies this branching and repair mindset to scientific artifact
management.

Long-term agent memory is a related but distinct layer. Generative Agents
maintains a memory stream with retrieval and reflection for interactive
behavior \cite{generative-agents}; MemGPT treats context management as an
explicit hierarchy between working memory and external storage
\cite{memgpt}; and MemoryBank studies retrieval and updating of long-term
memories for dialogue agents \cite{memorybank}. TeleMem extends this line to
long-term and multimodal memory with narrative extraction and structured
memory consolidation \cite{telemem}. These systems preserve useful context
across interactions, whereas XScientist preserves typed research transitions,
content-addressed evidence, failed branches, and authority-bound gates. A
memory store can help an agent remember what happened; it does not by itself
establish which recorded state is authoritative or whether a scientific claim
is replayable.

MLflow and DVC are representative tools for machine-learning lifecycle
tracking and data/model versioning \cite{mlflow,dvc}. They address important
parts of reproducibility. DataLad directly addresses the joint management of
code, data, and their relationships using versioned datasets
\cite{datalad}. Git's objects, trees, commits, and references provide an
established model for content identity and version history \cite{pro-git}.
XScientist builds on these ideas: its distinguishing design choices are the
typed research-state transitions, claim/evidence closure rules, and authority
conditions imposed on those records. File versioning, content addressing,
and preservation of execution provenance are not new contributions of this
paper. The experiment in Section~\ref{sec:evaluation} includes ordinary Git
and a whole-directory hash manifest as baselines for byte changes.

Snakemake and Nextflow provide mature approaches to executable computational
workflows \cite{snakemake,nextflow}. Sandve et al. recommend preserving exact
inputs and programs and connecting textual statements to underlying results
\cite{repro-computational}; the latter anticipates an important motivation for claim
anchors. The FAIR principles address findability, accessibility,
interoperability, and reuse of research data and metadata \cite{fair}.
These principles motivate the use of explicit provenance and portable
records, but an ARA export by itself does not establish FAIR compliance,
statistical validity, or independent reproducibility.

PROV-DM models entities, activities, and agents and their provenance relations
\cite{prov-dm}; RO-Crate packages research artifacts with machine-readable
metadata \cite{ro-crate}. Workflow Run RO-Crate extends this approach to
workflow executions, with profiles for inputs, outputs, code, and finer-grained
execution provenance across workflow systems \cite{workflow-run-ro-crate}.
Its reported applications also include re-execution and human-review records.
Portable run packaging, layered provenance, and recording review are therefore
established capabilities. ARA is a concrete format within XScientist, coupled
to its research transitions and policy gates; it is not proposed as the first
format for those broader capabilities. This implementation has no validated
mapping to Workflow Run RO-Crate, PROV-DM, or arbitrary third-party histories.
Cross-system interoperability remains to be demonstrated with independent
producers, consumers, and explicit preservation of policy semantics.

\subsection{Review, integrity, and benchmark comparison}
Reviewer-oriented systems such as DeepReviewer explore stronger automated
review signals \cite{deepreviewer}. Anti-Autoresearch motivates evidence-led
scrutiny of generated research artifacts \cite{anti-autoresearch}. XScientist
integrates reviewer-style repair with deterministic integrity forensics, but
keeps these checks conservative: they are safeguards and audit aids, not
proofs of scientific correctness. This distinction is especially important
for open review: an internal LLM review, an external AI report, and qualified
human certification are different forms of evidence and must not be collapsed
into one status.

MLE-bench evaluates machine-learning engineering agents on 75 Kaggle
competitions with executable grading, while ScienceAgentBench evaluates 102
data-driven scientific tasks drawn from 44 peer-reviewed publications
\cite{mle-bench,scienceagentbench}. Those benchmarks measure task-solving
ability and execution outcomes. Our study instead measures recorded-artifact
validation and replay on a human-specified workload; it does not establish
competitive task-solving or end-to-end discovery performance.

Several newer suites make a more defensible cross-system comparison possible,
provided that their native endpoints are kept separate. CORE-Bench evaluates
computational reproduction of 90 papers through 270 tasks at easy, medium, and
hard environment levels \cite{core-bench}; PaperBench evaluates replication of
20 ICML papers with 8,316 rubric leaves and a weighted replication score
\cite{paperbench}. AIRS-Bench provides 20 bounded research tasks with
task-level objective evaluators \cite{airs-bench}, while ResearchGym packages
39 subtasks from five recent papers into longer hypothesis--experiment loops
\cite{researchgym}. AstaBench provides a broader suite spanning literature,
code, data analysis, and end-to-end discovery, but its sub-benchmarks use
different programmatic and LLM-based judges \cite{astabench}. RE-Bench supplies
an eight-hour ML research-engineering track with a measured expert reference
\cite{re-bench}. These suites are complementary: none should be treated as a
universal autonomous-discovery score, and a benchmark result must not be
combined with the artifact-integrity results reported here.

The difficulty is not only that the suites use different tasks. Autonomous
research systems differ in model access, tool permissions, search depth,
human intervention, stopping rules, retry budgets, and the way failed or
incomplete runs enter the denominator. A reported paper-quality score,
task-success rate, wet-lab outcome, and artifact-audit signal therefore answer
different questions. A fair horizontal comparison must freeze the task and
data contract, model and budget, tool and network access, evaluator, and
failure accounting for every arm. We follow this rule in the bounded pilots
below and treat published results as contextual evidence rather than a pooled
ranking.

\section{The XScientist Protocol}

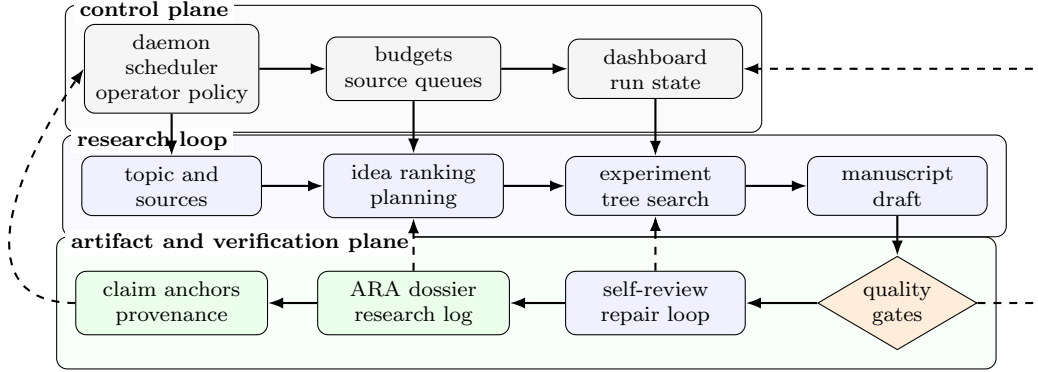
\begin{figure}[htbp]
\centering
\begin{tikzpicture}[
  x=1cm,
  y=1cm,
  every node/.style={font=\scriptsize},
  lane/.style={
    draw,
    rounded corners,
    inner sep=7pt,
    fill=#1,
    fill opacity=0.13,
    text opacity=1
  },
  lane label/.style={
    font=\scriptsize\bfseries,
    anchor=west,
    fill=white,
    inner xsep=2pt,
    inner ysep=1pt
  },
  block/.style={
    draw,
    rounded corners,
    align=center,
    text width=2.12cm,
    minimum height=0.72cm,
    fill=blue!6
  },
  artifact/.style={
    draw,
    rounded corners,
    align=center,
    text width=2.28cm,
    minimum height=0.72cm,
    fill=green!8
  },
  gate/.style={
    draw,
    diamond,
    aspect=1.7,
    align=center,
    inner sep=1.1pt,
    fill=orange!14
  },
  smallblock/.style={
    draw,
    rounded corners,
    align=center,
    text width=2.05cm,
    minimum height=0.66cm,
    fill=gray!8
  },
  arrow/.style={-{Latex[length=2mm]}, thick},
  feedback/.style={-{Latex[length=2mm]}, thick, dashed}
]
\node[smallblock] (daemon) at (0,1.55) {daemon scheduler\\operator policy};
\node[smallblock] (budget) at (3.2,1.55) {budgets\\source queues};
\node[smallblock] (dash) at (6.4,1.55) {dashboard\\run state};

\node[block] (sources) at (0,0) {topic and\\sources};
\node[block] (ideas) at (3.2,0) {idea ranking\\planning};
\node[block] (experiments) at (6.4,0) {experiment\\tree search};
\node[block] (writeup) at (9.6,0) {manuscript\\draft};

\node[gate] (gates) at (9.6,-1.55) {quality\\gates};
\node[block] (review) at (6.4,-1.55) {self-review\\repair loop};
\node[artifact] (ara) at (3.2,-1.55) {ARA dossier\\research log};
\node[artifact] (claims) at (0,-1.55) {claim anchors\\provenance};

\begin{scope}[on background layer]
  \node[lane=gray!20, fit=(daemon)(budget)(dash)] (controlplane) {};
  \node[lane=blue!18, fit=(sources)(ideas)(experiments)(writeup)] (researchloop) {};
  \node[lane=green!18, fit=(claims)(ara)(review)(gates)] (artifactplane) {};
\end{scope}
\node[lane label] at ([xshift=0.12cm,yshift=-0.08cm]controlplane.north west) {control plane};
\node[lane label] at ([xshift=0.12cm,yshift=-0.08cm]researchloop.north west) {research loop};
\node[lane label] at ([xshift=0.12cm,yshift=-0.08cm]artifactplane.north west) {artifact and verification plane};

\draw[arrow] (daemon) -- (budget);
\draw[arrow] (budget) -- (dash);
\draw[arrow] (daemon.south) -- (sources.north);
\draw[arrow] (budget.south) -- (ideas.north);
\draw[arrow] (dash.south) -- (experiments.north);

\draw[arrow] (sources) -- (ideas);
\draw[arrow] (ideas) -- (experiments);
\draw[arrow] (experiments) -- (writeup);
\draw[arrow] (writeup) -- (gates);
\draw[arrow] (gates) -- (review);
\draw[arrow] (review) -- (ara);
\draw[arrow] (ara) -- (claims);

\draw[feedback] (review.north) -- (experiments.south);
\draw[feedback] (ara.north) -- (ideas.south);
\draw[feedback] (claims.west) to[out=180,in=235,looseness=1.25] (daemon.west);
\draw[feedback] (gates.east) -- ++(0.82,0) |- (dash.east);
\end{tikzpicture}
\caption{XScientist research protocol flowchart. The upper control plane
schedules sources and budgets, the middle loop performs scientific work, and
the lower artifact plane records evidence, gates, repairs, and provenance for
later audit or forked continuation. Dashed edges denote long-horizon feedback.}
\label{fig:pipeline}
\end{figure}

Figure \ref{fig:pipeline} summarizes three logical planes. The control plane
budgets, schedules, and interrupts work. The research loop proposes experiments,
executes code, and drafts manuscripts. The artifact and verification plane
records evidence and decisions and checks the conditions for continuation or
promotion. These are logical responsibilities within a local application, not
a kernel or distributed operating system. A manuscript is a view over this
state, rather than its only source of truth.

\subsection{Execution and Artifact Model}

XScientist supports individual projects, batch generation, and long-running
daemon operation through a shared artifact model. Each mode writes structured
JSON and Markdown records of ideas, research plans, experiments, reviewer
findings, repairs, manuscript candidates, quality and integrity checks, source
provenance, and handoffs. These records provide a stable interface for agents
and humans to inspect and continue the work across execution modes.

The workflow begins by converting topics and source queues into ranked ideas
and plans that specify tasks, expected outputs, baselines, acceptance checks,
and execution policies. A tree-search-style experiment loop then records
code, stdout/stderr, metrics, plots, and failure states. Failed branches remain
available to explain subsequent repairs, softened claims, or decisions to
stop a branch. The writing layer produces manuscript candidates and structured
metadata; text and figure review identify issues and schedule repairs using
novelty, rigor, clarity, and reproducibility criteria. Management views expose
the resulting papers, unresolved issues, and repair priorities to operators.

Research VCS records this workflow as typed objects and checkpoints. A
checkpoint can add a question, hypothesis, experiment attempt, observation,
evidence item, claim, review, reproduction, or gate decision, together with
content hashes, actor and environment receipts, and explicit negative
outcomes. ARA is the portable export of one run, while Research VCS provides
the append-only history for comparing multiple runs, branches, and semantic
corrections. Mutable working state and refs remain separate from immutable
objects and checkpoints; corrections create new objects linked by
\texttt{supersedes} or use a typed revert checkpoint.

Execution is governed by explicit budgets and review boundaries. The policy
engine distinguishes local syntax and integrity checks from replay and
independent verification: passing a local check does not establish scientific
truth. A research cycle need not advance to a full manuscript. A failed sample
gate, a violated truth contract, a hard integrity flag, or an operator pause
can stop the run, and these stopping states are retained as valid outcomes.

\subsection{Typed History, Branches, and Corrections}

The Research VCS is the repository-level protocol that gives the paper's
``git-like'' analogy precise meaning. It is not a wrapper around ordinary Git
commands. Git may provide storage, but the scientific semantics are defined by
typed research objects, content-addressed identities, checkpoint-parent
relations, refs, and policy validation.

The object vocabulary covers the stages that reviewers need to inspect:
questions and search plans, source snapshots and passage evidence, hypotheses,
preregistrations, experiment attempts, observations, metrics, evidence,
claims, reviews, gate decisions, manuscripts, reproductions, and agent
evaluations. Objects carry a schema identifier, lifecycle state, typed
relations, actor and provenance fields, and a canonical content hash. Some
fields may be empty in lightweight records; stronger closure profiles require
the corresponding evidence and receipts.

An ordinary checkpoint can add objects or record a semantic correction, but it
cannot overwrite an existing object. A correction creates a new object linked
with \texttt{supersedes}; a rollback creates a checkpoint binding the commit
and checkpoint hash being reverted. Failed, timed-out, cancelled, negative,
contradictory, and rejected attempts, when recorded and committed, remain
inspectable in the reachable checkpoint history. This does not recover events
never recorded or preserve an unmerged branch whose only ref was removed.
Append-only semantics are enforced through the protocol API; direct backend
access remains within the operator's trust boundary.

\subsection{Evidence Closure and Authority Boundaries}

Schema validation constrains representation, while integrity checks test
consistency with declared identities. Separately, the closure auditor reports
\texttt{trace},
\texttt{replay}, and \texttt{verify}: trace checks the recorded claim/evidence
lineage; replay checks that required execution prerequisites are represented;
verify adds the required review, gate, and reproduction records. A
\texttt{replay\_ready} status is not evidence of a successful execution.
Actual re-execution produces a separate report, evaluated against an explicit
metric and tolerance.

A matching hash establishes consistency with the recorded content, not that
an event occurred or that a measurement is scientifically valid. The local
implementation can verify Ed25519 signatures and bind receipts to evidence;
establishing the external trust root, evaluator independence, and appropriate
domain expertise remains a deployment responsibility.

The publication-oriented profile requires confirmatory and reproduction
records to bind the relevant evidence, producers, evaluators, and attempt
dispositions. Other profiles and legacy records can be less complete. A claim
can be descriptive without
being causal or transferable; stronger statuses require stronger evidence and
independent gates. The implementation intentionally fails closed when a
required link is absent instead of inferring completion from prose or model
self-assessment.

\subsubsection{Portable History and Scientific Conflicts}

ARA and Research VCS solve adjacent problems. ARA is a portable export
for one end-to-end run, suitable for copying to an issue, archive, or another
agent. Research VCS is the durable history across runs, with named branches,
semantic diffs, merges, blame, policy decisions, and reproduction receipts.
The project runner projects ARA evidence and claim references into Research
VCS and binds the manifest and exploration-graph hashes. This is a specific
bridge, not a lossless import of arbitrary third-party research histories.
Projected claims remain drafts until the required gates are satisfied.

Two branches may contain byte-compatible files yet support incompatible
scientific interpretations. Research relations therefore distinguish
\texttt{supports}, \texttt{refutes}, \texttt{contradicts},
\texttt{depends\_on}, and \texttt{supersedes}. A merge records the selected
resolution policy; it cannot establish truth by choosing the last writer or
the largest metric. Comparable branches additionally need aligned task,
dataset split, metric, budget, and evaluator definitions. Preserving both
branches exposes disagreement but does not resolve it scientifically.

The checkpoint-parent DAG supplies causal ordering for the recorded history.
Timestamps are provenance metadata and do not establish that preregistration
preceded execution or that an evaluator was independent. Confirmatory
publication uses stronger state and authority bindings. This separation is
central to the Git analogy: a commit identifies a research transition, while
accepting its scientific interpretation remains an evidence-dependent act.

\subsection{Portable Artifacts and Continuation}

The run-level exchange interface is the Agent-Native Research
Artifact (ARA). An ARA is a directory rooted at a \texttt{manifest.json} file
and an \texttt{exploration\_graph.json} file. It is designed to be read by
another agent without requiring that agent to reverse-engineer the final
paper.

\subsubsection{Exploration DAG}

Each ARA stores an exploration DAG whose nodes represent exported experiment
states, including repair, failure, and ablation states when those states were
recorded by the producer. Edges encode the parent-to-child evolution of the
run. Manuscript candidates and related pipeline artifacts can be retained as
separate records; they are not necessarily experiment nodes. The validator
checks node identity, edge endpoints, counts, and acyclicity. The producer can
also write a summary and a self-contained HTML visualization.

The exploration DAG is intentionally not a polished success-only lineage. It
includes bug states, discarded branches, and repaired branches because those
states are often the most important for later review. A downstream agent that
sees only the final successful experiment cannot know whether the result was
robust, whether a metric was changed midstream, or whether a failed baseline
was silently dropped. ARA makes those transitions inspectable.

\begin{figure}[htbp]
\centering
\begin{tikzpicture}[
  x=1cm,
  y=1cm,
  every node/.style={font=\small},
  state/.style={
    draw,
    rounded corners,
    align=center,
    minimum width=2.05cm,
    minimum height=0.78cm,
    fill=gray!8
  },
  good/.style={state, fill=green!8},
  warn/.style={state, fill=orange!12},
  paper/.style={state, fill=blue!8},
  arrow/.style={-{Latex[length=2mm]}, thick}
]
\node[state] (root) at (0,0) {root\\question};
\node[good] (exp) at (2.8,0) {experiment\\node};
\node[warn] (fail) at (5.6,1.2) {failed\\branch};
\node[good] (repair) at (8.4,1.2) {repair\\rerun};
\node[good] (ablate) at (5.6,-1.2) {ablation\\node};
\node[paper] (candidate) at (8.4,-1.2) {manuscript\\candidate};
\node[state] (claim) at (11.2,-0.25) {claimref\\anchor};
\node[state] (fork) at (11.2,-1.95) {fork\\seed};

\draw[arrow] (root) -- (exp);
\draw[arrow] (exp) -- (fail);
\draw[arrow] (fail) -- (repair);
\draw[arrow] (repair) -- (candidate);
\draw[arrow] (exp) -- (ablate);
\draw[arrow] (ablate) -- (candidate);
\draw[arrow] (candidate) -- (claim);
\draw[arrow] (candidate) -- (fork);
\end{tikzpicture}
\caption{Conceptual science exploration DAG. ARA preserves recorded successful
and failed branches, allowing later agents to audit or fork a specific node.
The manuscript and claim records shown here are relations and derived views;
they need not be experiment nodes in every export.}
\label{fig:tree}
\end{figure}
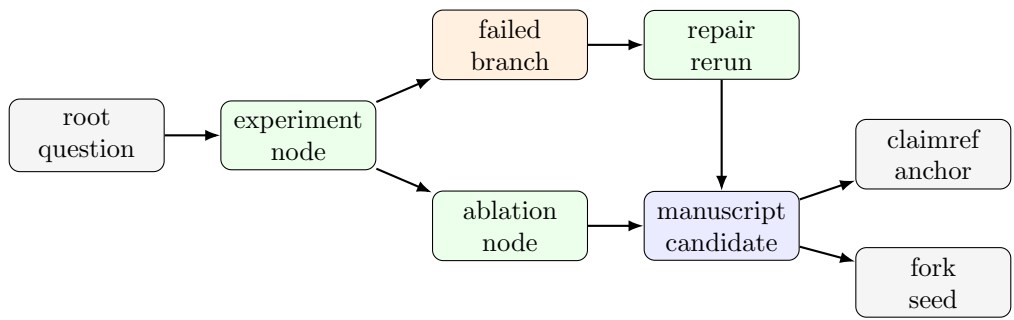

\subsubsection{Identity, Claims, and Operational Views}

ARA uses content hashes to identify node payloads and manifest revisions.
Current producers use the \texttt{ara.node-identity.v2} profile, which binds a
SHA-256 digest of the complete stripped UTF-8 source, normalized metric fields, and
declared extra references such as context or execution receipts. Consumers
retain legacy identity support rather than silently reinterpreting old
artifacts. Forked runs preserve parent pointers, so a child run can identify
the earlier ARA and node that seeded it. These hashes establish local content
integrity, not the truth of a measurement or the independence of its producer.
ARA node identities retain their own serialization profile for compatibility.
Newer trust-boundary commitments use the stricter
\texttt{xscientist.canonical-json.v1} profile, with Python and JavaScript
conformance fixtures. Those fixtures exercise that separate byte-level
boundary; they do not establish cross-language ARA node-identity conformance
or a second complete protocol implementation.

In practice, this gives XScientist a commit-like substrate. The final PDF is
only one rendered view. The underlying research state can be diffed, logged,
forked, or frozen. When an ARA is committed to a normal git repository, git
records file-level snapshots, while ARA commands expose node-level scientific
history inside those snapshots.

Generated manuscripts can include invisible \texttt{\textbackslash claimref}
markers after quantitative or evidence-sensitive claims. A claim registry
scans the source and writes the resolved or unresolved claim records under
the ARA. The marker is a pointer, not a judgment that the cited node entails
the claim. Coverage and resolution status can be summarized for quality gates
and downstream reviewers; missing or unresolved anchors remain explicit.

The installed CLI \texttt{xscientist ara} exposes operational verbs:
\texttt{inspect}, \texttt{exec}, \texttt{fork}, \texttt{freeze},
\texttt{validate}, \texttt{verify}, \texttt{diff}, \texttt{log}, and
\texttt{refs}. These commands let an operator inspect a node, re-execute its
code, create a new seed ARA, compare two ARAs, or walk the provenance chain.
The same graph data underlies the HTML visualization, the log view, and
node-level diffs.

\subsubsection{Portable Exchange}

The base \texttt{ara.v1} contract requires two files: \texttt{manifest.json}
and \texttt{exploration\_graph.json}. Optional but common files include per-node
\texttt{code.py}, \texttt{term\_out.log}, \texttt{metrics.json},
\texttt{plots.json}, environment snapshots, claim records, verification
reports, repair history, Pareto candidate pools, and a human-facing HTML graph
view. Legacy producers may declare absent optional artifacts in
\texttt{manifest.missing}. This does not waive the requirements of stronger
profiles: an authority-bound run also requires its attempt index, referenced
content-addressed object, and portable authority ledgers. Missing or invalid
required records cause validation to fail.

The complete directory contract is provided in Appendix~\ref{app:ara-contract};
the main text keeps the protocol invariant rather than a file inventory.

ARA uses ordinary files so that an export can be copied, archived, attached to
an issue, reviewed in a pull request, or consumed by another agent without
running an XScientist service. Databases and object stores remain useful
backends for querying or scale; the protocol keeps the portable boundary
independent of those services. As in software repositories, different tools
can provide richer views over the same explicitly versioned state.

Autonomous research systems require checks that are deterministic enough to
be run in CI and explicit enough to be audited later. XScientist implements
several such controls.

\subsection{Review, Repair, and Integrity Controls}

The self-review subsystem runs LLM-based and VLM-assisted reviews, extracts
structured issues, prioritizes repairs, and records repair attempts. The
system tracks whether reviewer objections were addressed, whether manuscript
changes regressed earlier claims, and whether quality thresholds were met.

Repair is represented as a trace rather than as a single overwrite. The
system records which issue motivated a repair, which artifact was modified,
which checks were rerun, and whether the repair introduced new risks. This is
important for generated manuscripts because a fluent rewrite can hide evidence
loss. XScientist therefore treats repair coverage and regression detection as
first-class quality signals.

\subsubsection{Quality and Integrity Gates}

Truth contracts translate a research plan into explicit constraints. They
separate objective facts, comparability constraints, artifact-binding rules,
branch-state rules, and value guardrails. For example, a manuscript should not
claim a baseline-comparable improvement unless the experiment registry
contains a completed run on the declared dataset, metric, and baseline.

The categories are intentionally broad: objective facts, comparability
constraints, artifact-binding rules, branch-state transitions, and value
guardrails. They are not a formal proof system. They are a pragmatic checklist
that converts ambiguous planner intent into concrete conditions that can be
checked later by gates, reviewers, or downstream agents.

Before full generation, sample gates can require that a small planned task
has a completed experiment record, a budget audit, passing acceptance checks,
and a result summary. This prevents the system from scaling a flawed plan into
a full paper when the cheap preliminary evidence is missing.

XScientist adapts an evidence-ledger pattern inspired by
Anti-Autoresearch into the manuscript stage. The integrity forensics pass is
not an authorship detector. It builds a span-anchored ledger from the
manuscript source, applies model-free consistency checks, and writes a
deterministic adjudication report. Hard findings can block submission-ready
status; soft findings are reported for reviewer attention.

The deterministic design is deliberate. Model-based critique is useful, but a
submission gate needs repeatable behavior. Integrity forensics therefore
focuses on checks that can be rerun with the same result: suspicious pipeline
artifacts in the manuscript, unsupported strong wording, arithmetic drift in
reported deltas, citation placeholders, repeated caption/table presentation
signals, and evidence-ledger gaps. It is a conservative screen for risks that
deserve attention before a paper is treated as submission-ready. Passing the
screen does not establish citation entailment, statistical validity, data
provenance, or scientific truth; false negatives and false positives remain
possible and require human adjudication.

Recent XScientist runs also write decision logs for workflow strategy,
sample-gate outcomes, and model-provider choices. Gate preconditions are
checked before expensive or high-risk stages proceed. For example, a quality
gate that requires high-quality mode should not run under a cheaper classic
pipeline by accident; a full-generation stage should not proceed when the
sample gate is blocked. These checks turn implicit workflow assumptions into
explicit runtime contracts.

\begin{table}[htbp]
\centering
\small
\begin{tabularx}{\linewidth}{@{}L{0.22\linewidth}L{0.32\linewidth}Y@{}}
\toprule
Control & Signal & Effect \\
\midrule
Sample gate & Sample task completion, budget audit, acceptance checks &
Blocks full generation when cheap preliminary evidence is absent. \\
Truth contract & Plan-derived facts, comparability rules, artifact bindings &
Checks registered facts and comparability conditions before promotion. \\
Integrity forensics & Span ledger, deterministic consistency checks & Raises
hard or soft manuscript risk flags. \\
Review repair trace & Structured issues, repair attempts, regression checks &
Shows how reviewer objections were addressed. \\
Gate preconditions & Workflow mode, quality mode, gate requirements & Stops
incompatible stage combinations before expensive execution. \\
\bottomrule
\end{tabularx}
\normalsize
\caption{Quality controls in XScientist. The goal is not to prove correctness
automatically, but to make risk states explicit and reproducible.}
\label{tab:quality-controls}
\end{table}

\section{Implementation and Operation}

The daemon mode treats research as an ongoing operation rather than a single
script invocation. It supports source queues, source health tracking,
dashboard serving, daily reports, handoff briefs, pause/resume controls,
source boosting and cooldown, failure protection, and strategy feedback.

Operational state accumulates across cycles. Source health, blocked gates,
remaining budget, and pending review actions inform subsequent scheduling.
The scheduler records these signals and control decisions. This capability
supports long-running operation; its effect on scientific productivity and
multi-day reliability is not evaluated in this paper.

\subsection{Supervised Operation}

The daemon is designed for supervised autonomy. Operators can pause and
resume, stop after the current cycle, force or boost a source, apply a
cooldown, inspect dashboard state, and generate handoff reports. This makes
the system suitable for long-running local operation where full automation is
not always desirable. A useful autonomous research system should make it easy
to interrupt, inspect, and redirect.

The daemon collects feedback at several levels: paper-level quality signals,
source-level health, strategy-level success rates, repair pressure, and
process alignment. These signals can influence later scheduling and execution
policy. For example, a source that repeatedly produces blocked sample gates
can be cooled down, while a source producing useful review-hardening tasks can
be boosted when a manuscript is near submission.

\subsection{Reproducibility and Release}

XScientist separates implementation code from a user-selected research
workspace. The output location is configuration-dependent and should be
recorded explicitly. For reproducibility, a report should cite the public
repository, configuration, model choices, output directory, and the ARA root
used for any generated paper. For the present system report, the canonical
repository is
\url{https://github.com/smileformylove/XScientist}.

The recommended release workflow for a generated manuscript is:
\begin{enumerate}[leftmargin=*]
  \item run the project or daemon under an explicit output root and budget;
  \item inspect the generated manuscript, ARA, review reports, and integrity
  forensics;
  \item verify or re-execute important ARA nodes where feasible;
  \item check claim coverage and manually review evidence-sensitive claims;
  \item package the manuscript and artifact references with the review archive.
\end{enumerate}

This workflow does not guarantee correctness, but it raises the cost of
silent evidence loss and makes later review more concrete.

\section{Experimental Evaluation}
\label{sec:evaluation}

The evaluation tests the paper's central claim at the artifact boundary. We ask
whether the protocol (i) preserves a replayable record from real computation,
(ii) distinguishes different classes of inconsistency, and (iii) gives a
reviewer more specific evidence than a changed-file signal. A fourth question
compares those outcomes with ordinary Git and complete byte hashing, including
serialization changes that preserve parsed JSON values. These are mechanism
questions, not a proxy for autonomous-discovery quality.

The study uses 20 real executions on the public Breast Cancer Wisconsin
dataset \cite{wdbc,sklearn}, producing 20 intact ARAs. Each is copied into six
fixed single-fault variants (changed code, changed metric, graph cycle,
dangling endpoint, missing code, and invalid schema type), yielding 120 paired
fixtures. The primary endpoints are per-fault rejection counts (out of 20
paired variants) and rejection of intact controls (out of 20); fault-class
coverage in Figure~\ref{fig:fault-coverage} and Table~\ref{tab:measured-faults}
is a derived summary requiring all 20 variants in a class to be rejected. A
later post hoc audit additionally counted whether saved reports contained a
non-empty reason and location; this records signal presence, not diagnostic
accuracy. The complete protocol, environment, hashes, and per-package reports
appear in the repository review archive.

The quantitative baselines are whole-directory SHA-256, plain Git, and a
normalized-JSON hash that canonicalizes JSON object order and whitespace. They
operate on the same saved packages and use an externally retained intact
reference, whereas XScientist's integrity check uses identities declared in
the candidate package. This difference in custody is part of the comparison.
For external task performance we report a common-schedule AIRS-Bench SICK
generated-program study, a matched ten-task ScienceAgentBench subset, a
20-item public-data LitQA2 scorer pilot from the AstaBench family, and a
CORE-Bench Hard preparation in which a five-capsule manifest was frozen but
only one exploratory capsule was executed. All are named as bounded
feasibility studies rather than full-suite or leaderboard results. Other public suites and
their unfulfilled hardware or data contracts are documented in Appendix~\ref{app:benchmarks}
\cite{airs-bench,researchgym,scienceagentbench,core-bench,paperbench,astabench,mle-bench,re-bench}.

\subsection{Results}

Cross-system scores are interpretable only under a frozen comparison contract.
The contract below makes the sources of heterogeneity explicit before any
native endpoint is displayed.

\begin{table}[htbp]
\centering
\small
\begin{tabularx}{\linewidth}{@{}L{0.22\linewidth}L{0.37\linewidth}Y@{}}
\toprule
Dimension & Frozen condition & Why it matters \\
\midrule
Task and data & Task revision, data hashes, public/held-out split & A changed split or prompt can change the scientific question. \\
Model and provider & Model identifier, endpoint, temperature, reasoning effort, seed & Provider behavior and sampling can dominate small score gaps. \\
Budget and tools & Token/call budget, wall-clock limit, retries, network and tool access & More search or broader tools can create an unreported advantage. \\
Evaluator and failures & Official evaluator version, output contract, denominator, exception policy & Dropped timeouts and malformed outputs make rates incomparable. \\
\bottomrule
\end{tabularx}
\caption{Minimum conditions for a defensible horizontal comparison. The
native task score and XScientist protocol endpoints remain separate even when
all four dimensions are frozen.}
\label{tab:comparison-contract}
\end{table}

The external benchmark matrix is a separately frozen endpoint. It keeps native
task performance separate from XScientist protocol evidence: the former is
scored by each benchmark's unchanged evaluator, while the latter records
artifact completeness, replay, closure, and retained failures. Table~\ref{tab:cross-system}
places the available autonomous-science systems on this evidence map. A native
score is comparable only when the task set, model, budget, seed policy, and
evaluator are shared; published results from different protocols are retained
as context rather than pooled into a leaderboard.

\begin{table}[htbp]
\centering
\scriptsize
\setlength{\tabcolsep}{2pt}
\renewcommand{\arraystretch}{1.08}
\begin{tabularx}{\linewidth}{@{}L{0.31\linewidth}L{0.39\linewidth}Y@{}}
\toprule
Benchmark / task (metric) & Comparator arm(s) & XScientist \\
\midrule
AIRS-Bench SICK classification (accuracy)~\cite{airs-bench} & AIDE $0.8127$; AI Scientist-v2 $0.8047$ & $0.8182$ \\
AIRS-Bench SICK similarity (Spearman)~\cite{airs-bench} & AIDE $0.7131$; AI Scientist-v2 $0.7219$ & $0.7055$ \\
ScienceAgentBench verified subset (official task success)~\cite{scienceagentbench} & AI Scientist-v2 MinimalAgent $1/10$ (4 submissions; 6 provider failures) & XScientist MinimalAgent $2/10$ (4 submissions; 6 provider failures) \\
AstaBench LitQA2 scorer pilot (accuracy; public LAB-Bench data)~\cite{astabench} & Direct GPT-5.6 $10/20$ (19/20 responses) & XScientist router GPT-5.6 $8/20$ \\
CORE-Bench Hard exploratory capsule (written accuracy)~\cite{core-bench} & $0/1$; script failed under xlrd/Python~3.13 & $0/1$; same capsule execution failure \\
\bottomrule
\end{tabularx}
\caption{Head-to-head and native endpoints. AIRS-Bench rows are a bounded common-schedule experiment over 4,906 held-out rows per task; all three arms used GPT-5.6, temperature 0, seed 42, low reasoning effort, native sandbox execution, three candidates, and the unchanged official grader. ScienceAgentBench uses 10 fixed tasks; its denominator retains provider failures and evaluator exceptions. The AstaBench-family row uses 20 public LAB-Bench LitQA2 items and the pinned scorer with GPT-5.6 at seed 0; AstaBench's gated split mapping was unavailable, so this is a non-canonical scorer pilot (the direct arm returned 19/20 responses and the router arm 20/20). The CORE-Bench row is a paired one-capsule retry: both arms reached the evaluator but received $0/1$ after the capsule's \texttt{script.py} failed with the \texttt{xlrd~1.2.0}/Python~3.13 incompatibility; this is a reproducibility-failure record, not a successful scientific-task result. A prior protocol-only $1/1$ trajectory is retained separately and is not pooled with this retry. Artifact-integrity results are reported separately in the protocol study.}
\label{tab:cross-system}
\end{table}

The AIRS rows provide the strongest direct comparison in this revision. On
classification, XScientist scores $0.8182$, compared with $0.8127$ for AIDE
and $0.8047$ for AI Scientist-v2; on similarity, XScientist scores $0.7055$,
below $0.7131$ for AIDE and $0.7219$ for AI Scientist-v2. These are official held-out scores, not
validation estimates, but they come from an adapted CPU schedule and a single
fixed seed. They demonstrate a common executable comparison and delimit the
claim: XScientist's contribution is the research-state and audit layer, not a
measured accuracy win on this pilot.
Candidate failures are retained in the archive. This is a bounded replication
rather than a preregistered leaderboard entry. The separate ScienceAgentBench pilot
uses the same ten-task data contract and official evaluator but different
native prompts at the MinimalAgent layer; it gives 1/10 for AI Scientist-v2
and 2/10 for XScientist. Both arms also contain provider or generation
exceptions, and only one task is a common formally returned task, so the raw
difference is not evidence of a method advantage. Both denominator counts
retain the evaluator exception; this is secondary evidence, not a protocol
ranking.

The AstaBench-family pilot is deliberately reported with a narrower claim. It
uses the public LAB-Bench LitQA2 data and the pinned LitQA2 scorer, with the
same 20 items, GPT-5.6, and seed 0 for a direct arm and the XScientist router.
The direct arm scores $10/20$ (19 responses) and the router scores $8/20$;
because the official AstaBench split mapping is gated, these numbers are a
non-canonical pilot and do not establish a benchmark advantage or disadvantage.

The CORE-Bench retry is reported as a failure diagnosis rather than a task
success claim. On the same capsule and provider contract, both arms reached the
official evaluator and scored $0/1$. The capsule's \texttt{script.py} failed
under \texttt{xlrd~1.2.0} on Python~3.13 because it calls the removed
\texttt{ElementTree.getiterator} method; the XScientist arm additionally
recorded one protocol step before submitting an unverified answer. The earlier
single-arm $1/1$ trajectory is retained as a separate exploratory record.

The published systems provide broader context but report different endpoints.
Table~\ref{tab:published-systems} records their denominators and metrics
without averaging them with the native AIRS scores.

\begin{table}[!tb]
\centering
\scriptsize
\setlength{\tabcolsep}{2pt}
\renewcommand{\arraystretch}{1.05}
\begin{tabularx}{\linewidth}{@{}L{0.18\linewidth}L{0.28\linewidth}Y@{}}
\toprule
System & Published task / benchmark & Native reported endpoint \\
\midrule
AI Scientist~\cite{ai-scientist} & Diffusion-paper generation; 51 ideas & 38 completed papers; mean automated reviewer score $3.82/6$. \emph{Paper-generation endpoint.} \\
AI Scientist-v2~\cite{ai-scientist-v2} & ICBINB workshop; 43 submissions & 3 selected manuscripts, one acceptance decision; accepted-paper mean $6.33/10$. \emph{Not an unbiased $1/3$ success rate.} \\
AIDE~\cite{aide,mle-bench} & MLE-bench (75); Weco-Kaggle Lite (16) & Any-medal pass@1 $16.9\pm1.1\%$ (mean$\pm$SE) with o1-preview; above human median on 8/16 Weco-Kaggle Lite tasks. \emph{ML-engineering endpoint.} \\
Agent Laboratory~\cite{agent-laboratory} & Autonomous study; 15 papers & Human experimental-quality means $2.6/5$, $3.2/5$, and $2.9/5$ for gpt-4o, o1-mini, and o1-preview. \emph{Human-judgment endpoint.} \\
AgentRxiv~\cite{agentrxiv} & MATH-500; sequential and parallel laboratories & Accuracy: $70.2\%$ baseline; best sequential $78.2\%$, parallel $79.8\%$. \emph{Mathematical-task endpoint.} \\
Kosmos~\cite{kosmos} & Factuality audit; 102 report statements & $79.4\%$ statement accuracy. \emph{Claim-quality endpoint.} \\
Co-Scientist~\cite{co-scientist} & AML candidate screening; five candidates & Preliminary inhibition for 3/5 candidates. \emph{Wet-lab endpoint.} \\
ScientistOne~\cite{scientistone} & Chain-of-evidence audit; 75 papers & Score verification $12/12$; method--code alignment $14/15$; references $0/337$. \emph{Evidence-audit endpoint.} \\
AutoResearch\-Claw~\cite{autoresearchclaw} & ARC-Bench; 25 topics & CoPilot $0.648$; Full-Auto $0.596$; AIDE-ML $0.511$; AI Scientist-v2 $0.419$. \emph{Rubric score; human mode differs.} \\
DeepScientist~\cite{deepscientist} & Month-scale discovery campaign & $>20{,}000$ GPU-hours; $\sim5{,}000$ ideas; $\sim1{,}100$ validated; 21 innovations. \emph{Campaign totals, not pass@1.} \\
AI-Researcher~\cite{ai-researcher} & Scientist-Bench; 22 papers & 93.8\% implementation completeness; correctness $2.65/5$. \emph{Judge- and level-dependent.} \\
FAR~\cite{far} & Combinatorics discovery funnel & 5,245 papers $\rightarrow$ 6,453 candidates $\rightarrow$ 77 expert-review items. \emph{Triage funnel, not solved accuracy.} \\
MARS~\cite{mars} & MLE-bench; 75 competitions & Above-median $65.8\%$; any-medal $56.0\%$; MARS+ any-medal $62.7\%$. \emph{Published controlled setting; A100/24 h.} \\
\bottomrule
\end{tabularx}
\caption{Published autonomous-science results used as context. The table reports each system's native endpoint in the outcome column; the italicized notes state why that endpoint cannot be pooled with the others or treated as a head-to-head score against XScientist.}
\label{tab:published-systems}
\end{table}

\begin{figure}[htbp]
\centering
\begin{tikzpicture}[
  x=0.82cm,
  y=0.62cm,
  every node/.style={font=\scriptsize},
  axis/.style={draw=black!55,line width=0.4pt},
  grid/.style={draw=black!15,line width=0.3pt},
  bar/.style={draw=black!25,line width=0.3pt}
]
\foreach \x in {0,1,2,3,4,5,6}
  \draw[grid] (\x,0.55) -- (\x,5.55);
\draw[axis] (0,0.55) -- (6.2,0.55);
\foreach \x/\lab in {0/0,1/1,2/2,3/3,4/4,5/5,6/6} {
  \draw[axis] (\x,0.55) -- (\x,0.43);
  \node[below] at (\x,0.39) {\lab};
}
\node[anchor=west] at (0,6.0) {predefined fault classes rejected (of 6)};
\node[anchor=east,align=right] at (-0.18,5) {Schema only};
\node[anchor=east,align=right] at (-0.18,4) {Conformance};
\node[anchor=east,align=right] at (-0.18,3) {Integrity};
\node[anchor=east,align=right] at (-0.18,2) {Replay};
\node[anchor=east,align=right] at (-0.18,1) {Combined};
\draw[bar,fill=blue!35] (0,4.78) rectangle (1,5.22);
\draw[bar,fill=teal!55] (0,3.78) rectangle (4,4.22);
\draw[bar,fill=cyan!45] (0,2.78) rectangle (3,3.22);
\draw[bar,fill=violet!45] (0,1.78) rectangle (2,2.22);
\draw[bar,fill=orange!75] (0,0.78) rectangle (6,1.22);
\node[anchor=west] at (1.12,5) {1/6};
\node[anchor=west] at (4.12,4) {4/6};
\node[anchor=west] at (3.12,3) {3/6};
\node[anchor=west] at (2.12,2) {2/6};
\node[anchor=west] at (6.08,1) {6/6};
\end{tikzpicture}
\caption{Matched coverage of the six predefined artifact-fault classes. Each bar
counts classes for which all 20 paired variants were rejected by that check;
all five checks accepted the 20 intact controls. The plot summarizes the
coverage matrix in Table~\ref{tab:measured-faults}; it is not a sensitivity
estimate for naturally occurring faults.}
\label{fig:fault-coverage}
\end{figure}
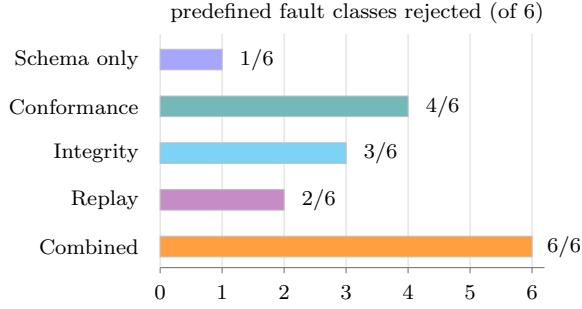

% Generated from raw summary.json by run_evaluation.py; do not edit counts.
\begin{table}[t]
\centering\scriptsize
\begin{tabular}{lcccccc}
\toprule
Check & Code & Metric & Cycle & Endpoint & No code & Type \\
\midrule
Schema Only & 0/20 & 0/20 & 0/20 & 0/20 & 0/20 & 20/20 \\
Conformance & 0/20 & 0/20 & 20/20 & 20/20 & 20/20 & 20/20 \\
Integrity & 20/20 & 20/20 & 0/20 & 0/20 & 20/20 & 0/20 \\
Replay & 0/20 & 20/20 & 0/20 & 0/20 & 20/20 & 0/20 \\
Combined & 20/20 & 20/20 & 20/20 & 20/20 & 20/20 & 20/20 \\
\bottomrule
\end{tabular}
\caption{Primary per-fault endpoint for the paired artifact-fault matrix. Each entry is the number of rejected packages out of 20 paired copies for that fixed fault class; the six columns are code tampering, metric tampering, a graph cycle, a dangling endpoint, missing code, and an invalid schema type. Every check accepted all 20 intact controls. The six classes are an exhaustive targeted fixture set, not independent samples or an estimate of natural-world sensitivity.}
\label{tab:measured-faults}
\end{table}

All 20 original executions produce valid packages and pass local re-execution
at the prespecified tolerance. Figure~\ref{fig:fault-coverage} and
Table~\ref{tab:measured-faults} show coverage of 1, 4, 3, 2, and 6 of the six
predefined fault classes for schema, structure, integrity, replay, and the
combined protocol, respectively. Every check accepts the intact controls. The
structure--integrity conjunction already covers all six selected classes;
replay adds executable evidence but no additional rejection on this targeted
set. These are paired fixture outcomes, not estimates of natural fault
prevalence or scientific validity.

Two misses clarify the boundaries. Appending an extra print statement changes
the source bytes but leaves balanced accuracy unchanged, so metric replay
accepts it and the content check rejects it. Conversely, adding a graph cycle
or dangling endpoint leaves the selected node's code and metric unchanged;
the tested node-hash/manifest-lock check and single-node replay accept these
packages, whereas structural validation rejects them. This is why none of
these checks alone is an adequate proxy for the others.

\subsection{Additional Storage Baselines and Representation Controls}

The storage comparison asks whether this coverage is simply a consequence of
checking bytes. Whole-directory SHA-256 and plain Git match the combined
protocol on the six selected fault classes and on the intact controls. A
normalized-JSON hash also accepts the representation-preserving rewrites,
whereas byte-level baselines flag them. Thus the experiment does not show a
detection advantage for XScientist or a unique serialization invariant. The
measured difference is semantic: protocol reports retain the violated
relation, typed field, or metric identity (for example, a self-loop at
\texttt{edges[0]}), while storage comparators report changed paths. The
Appendix~\ref{app:comparators} contains the exact comparator rules and
Appendix~\ref{app:diagnostics} contains diagnostic examples; the
signal-presence count is not a claim of root-cause accuracy or reviewer
benefit.

\subsection{Interpretation and Scope}

The computational workload establishes that the packages contain metrics
computed from real predictions and reproducible commands. The relevant systems
result is the combination of fault-class coverage, intact-control acceptance,
and protocol diagnostic context. Schema checks constrain representation; graph
checks add relational invariants; hashes check consistency with declared
identities or reference bytes; replay compares a declared metric under a
recorded command. A combined pass on a valid control is not an independent
reproduction or a scientific endorsement.

The workloads do not exercise autonomous hypothesis generation, multi-day
daemon reliability, or human review quality, and the native agent comparisons
are bounded CPU pilots rather than full discovery evaluations against
AI-Scientist-v2 or AIDE. They also do not measure the rate of naturally
occurring corruption. Those questions require matched tasks, budgets,
prospective runs, independent evaluators, and domain-appropriate outcomes.
The bundled demonstration's prewritten retrieval rates are synthetic examples
and are excluded from every empirical table in this paper.

\section{Discussion and Limitations}

XScientist does not make autonomous research automatically correct. The
system can organize evidence, preserve provenance, and expose warnings, but
human review remains necessary for scientific validity, ethical judgment, and
final submission decisions. The current system also has several engineering
limitations:

\begin{itemize}[leftmargin=*]
  \item Re-execution can be expensive or environment-dependent when nodes
  call external APIs, GPUs, or web services.
  \item Integrity forensics is deterministic but incomplete; its false-positive
  and false-negative rates on naturally occurring manuscripts have not been
  measured in this study.
  \item ARA is a protocol surface rather than a universal standard; broader
  adoption requires independent producers and consumers.
  \item The daemon provides operational controls, but no multi-day endurance
  or crash-recovery comparison is reported here. Long-running autonomy still
  needs explicit budgets, stop conditions, and operator oversight.
  \item The present paper is a system report. It documents the implementation
  and protocol surface with a narrow computational evaluation. It does not
  provide a controlled user study, human-versus-agent comparison, or a broad
  external discovery benchmark.
\end{itemize}

\subsection{Internal and External Validity}
The evaluation uses one small, well-understood dataset, simple classical
models, one platform, and author-designed faults. Multiple corrupted copies
of a package are paired test cases, not independent scientific studies.
The protocol and checker were developed in the same project, and the local
execution environment is not an independent verifier. The evaluated commit
is fixed; later changes may alter results and require a new evaluation.
The external task study is intentionally bounded. The AIRS rows are exact
official-evaluator scores on a 4,906-row held-out SICK generated-program task,
but use an adapted three-candidate CPU schedule and one seed rather than the
canonical AIRS agent protocol. ScienceAgentBench uses a fixed ten-task verified
subset and reports $1/10$ for AI Scientist-v2 and $2/10$ for XScientist; the
shared evaluator exception is retained in both denominators. A paired
CORE-Bench retry on capsule-9052293 reached the evaluator for both arms and
scored $0/1$ in each, because the capsule script failed under the pinned
xlrd/Python~3.13 environment. This is a reproducibility-failure record, not a
successful task result; the earlier single-arm $1/1$ trajectory is retained
separately. A valid full-suite
comparison must use the same agent-facing task, model, budget, seed/retry
policy, hardware, and evaluator for every system; native scores remain
separate from artifact and recovery metrics.

\subsection{Adversary Model}
The fault experiment tests inconsistencies against recorded identities. An
operator who controls source, data, records, and trust configuration can
construct a mutually consistent but fabricated history, including recomputing
hashes after changing the contents. Local content addressing cannot prevent
this. External attestations can bind a trusted verifier's statement to exact
bytes, but cannot repair an invalid study design or confer independence on a
colluding evaluator. These risks are outside the demonstrated detection
results.

\subsection{Ethics and Operational Safety}

Autonomous research tooling can create plausible but incorrect scientific
artifacts at high speed. XScientist is therefore designed to preserve
uncertainty, not to erase it. The system records failed branches, exposes
blocked gates, and explicitly states that integrity forensics is incomplete.
Users should not submit generated papers without human review, should disclose
substantial automated assistance when appropriate, and should avoid using the
system for domains where erroneous claims can cause direct harm without
additional expert oversight.

There is also an operational safety concern: long-running agents can consume
API credits, compute, and storage unexpectedly. XScientist includes output
isolation, preflight checks, daemon controls, and budget-oriented gates, but
operators remain responsible for configuring limits and monitoring runs.

\subsection{Implications}

The central design claim of XScientist is that a manuscript is a rendered view
of a research history. When runs branch, fail, and resume, the reviewable unit
is the history: the state transitions, evidence links, and decisions that
justify the current view. The controlled study supports the narrower systems
claim behind this design. XScientist matches storage baselines on the selected
fault coverage, while reporting reasons and diagnostic context tied to the
protocol. The designed continuation semantics extend this representation; their
effect on recovery cost still requires a longitudinal experiment.

The same design suggests a path toward collaborative autonomous science:
systems can exchange ARAs, fork one another's nodes, inspect claim/evidence
links without parsing PDFs, and use git-like diffs to review scientific
changes. This is an interoperability objective that still needs independent
producers and consumers. For human review, the immediate benefit is more
concrete: a reviewer can request an exact artifact or disputed transition,
and an author response can point to a new immutable record. A platform's
human-certification decision remains external to XScientist's internal gates.

\section{Conclusion}

XScientist implements a Git-like protocol for maintaining scientific state
across research runs. Research VCS records typed objects and transitions;
ARA exports run-level evidence and execution artifacts; closure and authority
checks constrain how those records can support a claim. The controlled
evaluation demonstrates complementary detection behavior across schema,
structure, content-integrity, and replay checks on real computational
artifacts. Complete hash comparison, normalized JSON hashing, and ordinary Git
match its fault coverage; the additional JSON controls distinguish byte
identity from the tested protocol invariants. The protocol reports invariant-
specific reasons and locations for the selected faults, while the study does
not establish general autonomous discovery performance.
The protocol provides an inspectable basis for continuation and review while
leaving scientific validity and independent certification explicit.

\section*{Availability}

The primary artifact for this report is the public XScientist GitHub
repository:
\begin{center}
\url{https://github.com/smileformylove/XScientist}
\end{center}
It contains the implementation, ARA protocol schemas, command-line tools,
examples, documentation, and tests. The arXiv source for this system report is
maintained in the same repository under:
\begin{center}
\texttt{paper/xscientist\_arxiv/}
\end{center}
The accompanying review archive contains the evaluation protocol, driver,
raw predictions, validation reports, generated tables, and complete evaluated
source. Saved-result verification can use that archive without fetching the
source from the remote repository. The local environment and exact dependency
versions are recorded with the run. Neither the earlier preprint nor this revision is represented as having
received human certification.

\section*{AI Assistance and Author Responsibility}

An OpenAI Codex coding assistant supported literature and source-code checks,
draft revision, preparation of the evaluation scripts, and local execution.
The computational workload and validation checks call no language-model
provider. AI-assisted drafting and internal automated checks are not
independent peer review. The named author remains responsible for the
manuscript, citations, interpretation of the recorded results, and the
declarations accompanying submission.

\section*{Acknowledgements}

XScientist incorporates modified AI-Scientist-v2 code and transitive AIDE
components, as documented in \texttt{THIRD\_PARTY\_NOTICES.md}. The audited
comparison snapshots and retained Apache-2.0 and MIT notices are provided in
the repository; the audit does not establish an exact historical acquisition
date. We also acknowledge the TeleMem project as related work on long-term
and multimodal memory for agentic systems. Other design references include
autoresearch, DeepReviewer, and Anti-Autoresearch. Their published benchmark
results are not inherited by XScientist.

\clearpage
\appendix

\section{Evaluation Contract}
\label{app:contract}

This appendix is part of the submitted PDF and records the material needed to
interpret the mechanism study without turning the main text into an execution
log. The frozen evaluation uses 20 real computational executions on the
Breast Cancer Wisconsin (Diagnostic) data, with ten fixed stratified splits and
two fixed models per split. Each execution exports one intact ARA. Six single
fault operators are applied independently to each ARA: source-code tampering,
metric tampering, a graph self-loop, a dangling edge endpoint, removal of the
replay source, and an invalid schema type. The resulting 120 variants are
paired with their 20 intact controls.

For a check, ``rejection'' means that the check rejects the predefined fixture
variant; it is not a sensitivity, precision, or security estimate. A false
rejection is rejection of an unchanged control. The replay check uses a
60-second timeout and an absolute balanced-accuracy tolerance of
$10^{-12}$. An exception is recorded as a failed check and retained with its
diagnostic. The six fault classes and their intended invariant are:

\begin{center}
\small
\begin{tabular}{ll}
\toprule
Fault operator & Intended invariant under test \\
\midrule
\texttt{code\_tamper} & source bytes agree with the declared node identity \\
\texttt{metric\_tamper} & recorded metric agrees with the declared identity and replay \\
\texttt{graph\_cycle} & exploration graph remains acyclic \\
\texttt{missing\_endpoint} & every edge endpoint names a present node \\
\texttt{missing\_code} & replay inputs required by the node are present \\
\texttt{schema\_type} & schema field types remain valid \\
\bottomrule
\end{tabular}
\end{center}

The complete protocol, source and data hashes, environment record, raw
packages, per-package reports, and generated tables are retained under
\texttt{paper/xscientist\_arxiv/evaluation/}. The primary table in the main
text reports class coverage so that the denominator is visible without
suggesting that the paired fixtures are independent observations.

\section{ARA Directory Contract}
\label{app:ara-contract}

The base \texttt{ara.v1} export requires the two entry-point files below.
Stronger conformance and authority profiles add mandatory records. The
file-based boundary can be archived, diffed, copied, or consumed by tools that
do not run an XScientist service.

\begin{table}[htbp]
\centering
\small
\begin{tabularx}{\linewidth}{@{}L{0.25\linewidth}L{0.14\linewidth}Y@{}}
\toprule
Artifact & Status & Purpose \\
\midrule
\texttt{manifest.json} & required & Entry point, schema version, counts,
references, provenance, and missing-artifact declarations. \\
\texttt{exploration\_graph.json} & required & Node and edge set for the
science exploration DAG. \\
\texttt{nodes/<id>/} & optional & Per-node code, metrics, logs, plots,
environment descriptors, and rerun script. \\
\texttt{claims/} & optional & Claim-to-node anchors extracted from manuscript
source. \\
\texttt{verify/} & optional & Re-execution reports and metric-drift checks. \\
\texttt{pipeline/} & optional & Mirrored planning, review, repair, and gate
artifacts for downstream agents. \\
\bottomrule
\end{tabularx}
\caption{Base ARA directory contract. Optional artifacts may be declared in
\texttt{manifest.missing}; this does not waive requirements of stronger
profiles.}
\label{tab:ara-contract}
\end{table}

\section{Comparator Definitions}
\label{app:comparators}

The storage baselines use an externally retained intact package. Whole-tree
SHA-256 compares each regular file's relative path, size, and raw-content
digest. Plain Git commits the intact package in an isolated repository and
checks tracked and untracked differences with hooks, signing, text
conversion, global attributes, and external configuration disabled. Neither
baseline interprets scientific relations.

The normalized-JSON comparator parses JSON with duplicate keys and non-finite
constants rejected, serializes object keys in sorted order with fixed
separators, and hashes the canonical bytes; non-JSON files remain byte-hashed.
It is used only for the representation controls and the same saved fault
packages. The XScientist C+I row in the audit is a conjunction of saved
conformance and integrity outcomes; it is not a new replay run. On the
selected faults, all three storage baselines and C+I reject 120/120 variants
and accept 20/20 intact controls. The normalized comparator accepts 20/20
representation controls; byte baselines flag those changes. These results
establish matched behavior for this fixture set, not general superiority.

\section{Diagnostic Examples}
\label{app:diagnostics}

The protocol report preserves an invariant-specific reason and a location in
addition to the changed path. Representative saved signals are:

\begin{center}
\footnotesize
\begin{tabular}{@{}L{0.17\linewidth}L{0.48\linewidth}L{0.27\linewidth}@{}}
\toprule
Fault & Representative protocol signal & Location \\
\midrule
Code change & code differs from the declared node identity & \path{nodes/experiment/code.py} \\
Metric change & metric differs from the declared node identity & \path{nodes/experiment/metrics.json} \\
Self-loop & node \texttt{experiment} links to itself & \texttt{edges[0]} \\
Missing endpoint & edge child \texttt{absent\_node} is not present & \texttt{edges[0]} \\
Missing code & replay requested but the source artifact is absent & \texttt{conformance} \\
Invalid type & \texttt{counts.nodes} is not an integer & \texttt{counts.nodes} \\
\bottomrule
\end{tabular}
\end{center}

Each signal is present for all 20 paired instances of its class in the saved
reports. This is a signal-presence endpoint, not a claim of root-cause
accuracy, reviewer-time reduction, or independent custody. A self-consistent
fabricated history can evade local checks if an operator is allowed to rewrite
the declared identities and trust configuration.

\section{External Benchmark Readiness}
\label{app:benchmarks}

The public suites below provide possible matched task endpoints. They are
listed to define the reproducible comparison contract. The per-suite source
pins, blocked conditions, and adapter commands are retained here so that the
main-table entries can be audited without duplicating the matrix.

The archived CORE-Bench preparation uses a temporary local path adaptation and
no Docker. The five-capsule manifest is a frozen task-selection record. A
paired retry on capsule-9052293 reached the official evaluator for both arms and
returned $0/1$ for each after the capsule script failed under xlrd/Python~3.13;
the earlier single-arm $1/1$ trajectory is retained separately and is not a
paired or leaderboard score. The AIRS adapter pins the official evaluator and uses a
separate evaluator directory/process for test-label injection; this layout is
not an OS or container security boundary. Candidate selection may read only the public training and
validation splits; the agent-facing test split remains unlabeled until the
official grader runs after submission. A canonical comparison must fix the
benchmark commit and data hashes, model/API identifier, prompts, seed and
retry policy, tool and network access, wall-clock and token budget, hardware,
output contract, stopping rule, and evaluator version. It should publish the
benchmark's native score beside protocol endpoints such as artifact
completeness, claim--evidence closure, replay success, failed-branch
retention, and recovery time, without averaging the two families.

The completed bounded runs are retained with their raw inputs and evaluator
outputs. The three-arm AIRS SICK study is under
\path{external_benchmarks/airs_agent/runs/airs_sick_generated_fair_20260922/};
its held-out classification scores are 0.8127 (AIDE), 0.8047
(AI Scientist-v2), and 0.8182 (XScientist), while
similarity scores are 0.7131, 0.7219, and 0.7055. The ScienceAgentBench comparison is under
\path{external_benchmarks/native_science/runs/sab_v2_full/} and
\path{external_benchmarks/native_science/runs/sab_xsci_full/}; its official
task-success counts are 1/10 and 2/10. The CORE-Bench preparation and
exploratory runs are under
\path{external_benchmarks/corebench_matched/}; the paired retry record reports
$0/1$ for both arms after the capsule's xlrd/Python~3.13 execution failure, and
the earlier protocol-only record reports $1/1$ separately. Neither is a
full-suite leaderboard result.
The AstaBench-family LitQA2 scorer pilot is under
\path{external_benchmarks/api_cpu_secondary/runs/20260921_190219_litqa2_20matched/};
on 20 public LAB-Bench items, the direct GPT-5.6 arm scored $10/20$ and the
XScientist router scored $8/20$. The receipt marks this result non-canonical
because the official AstaBench split mapping is gated.
These records are auditable feasibility evidence, not full-suite leaderboard
submissions.

\section{Reproduction Commands}
\label{app:commands}

From the repository root of a clean checkout, the saved-result verifier can be
run as follows:

\begin{verbatim}
python paper/xscientist_arxiv/evaluation/verify_results.py \
  --run paper/xscientist_arxiv/evaluation/raw/run_20260919_01 \
  --repo . \
  --report verification-local.json
\end{verbatim}

From the root of the extracted review bundle, use the bundled evaluated
source archive:

\begin{verbatim}
python manuscript/evaluation/verify_results.py \
  --run manuscript/evaluation/raw/run_20260919_01 \
  --source-archive source/xscientist-evaluated-source.tar \
  --report verification-local.json

python manuscript/evaluation/comparison_20260919/verify_comparison.py \
  --base-run manuscript/evaluation/raw/run_20260919_01 \
  --comparison-run manuscript/evaluation/comparison_20260919/run_20260919_03 \
  --source-archive source/xscientist-evaluated-source.tar
\end{verbatim}

These commands verify saved evidence and hashes; they do not claim external
custody or independent scientific review. Re-running the full computation
requires the pinned source snapshot and dependencies described in the
evaluation archive. The benchmark adapters similarly record source pins,
submission snapshots, grader stdout/stderr, and failure states, but an adapter
smoke test is not a benchmark score.

\end{document}